\documentclass[aps,preprintnumbers,amsmath,superscriptaddress,showpacs,twocolumn]{revtex4}

\usepackage{epsfig}
\usepackage{psfrag}
\usepackage{amsfonts}
\usepackage{graphicx}
\usepackage{dcolumn}
\usepackage{bm}
\newcommand{\be}{\begin{equation}}
\newcommand{\ee}{\end{equation}}
\newcommand{\ba}{\begin{eqnarray}}
\newcommand{\ea}{\end{eqnarray}}
\def\ie{ \emph{i.e., } }

\usepackage{CJK}

\begin{document}
\begin{CJK*}{GB}{gbsn}
\title{Color-electric conductivity in a viscous  quark-gluon plasma}

\author{Bing-feng Jiang(½¯±ù·å)}
\email{jiangbf@mails.ccnu.edu.cn} \affiliation{Center for Theoretical Physics and Department of Physics, Hubei Minzu University, Enshi, Hubei 445000,
China }

\author{Shao-wu Shi(ʯÉÜÎé)}
\affiliation{Center for Theoretical Physics and Department of Physics, Hubei Minzu University, Enshi, Hubei 445000,
China }

\author{De-fu Hou(ºîµÂ¸»)}
\email{houdf@mail.ccnu.edu.cn}\affiliation{Key Laboratory of Quark and Lepton Physics
(MOE) and Institute
of Particle Physics, Central China Normal
University, Wuhan, Hubei 430079, China}

\author{Jia-rong Li(Àî¼ÒÈÙ)}
\email{ljr@mail.ccnu.edu.cn} \affiliation{Key Laboratory of Quark and Lepton Physics
(MOE) and Institute
of Particle Physics, Central China Normal
University, Wuhan, Hubei 430079, China}

\date{\today}
\begin{abstract}
Several different transport processes, such as heat transport, momentum transport and charge transport, may take place at the same time in a thermal plasma system.  The corresponding transport coefficients are heat conductivity, shear viscosity and electric conductivity respectively. In the present paper, we will study the color-electric conductivity of the quark-gluon plasma (QGP) in presence of shear viscosity, which is focused on the connection between the charge transport and the momentum transport.  To achieve that goal, we solve the viscous chromohydrodynamic equations which are obtained from the QGP kinetic theory associated with the distribution function modified by shear viscosity. According to the solved color fluctuations of hydrodynamic quantities we obtain the induced color current through which the color-electric conductivity  is derived.
Numerical analysis show that
the conductivity properties  of the QGP are mainly demonstrated  by the longitudinal part of the color-electric conductivity. Shear viscosity has an appreciable impact on real and imaginary parts of the color-electric conductivity in some frequency region.
\end{abstract}

\maketitle
\section{Introduction}

People  have expected to find quark-gluon plasma (QGP), a special state of matter, by ultrarelativistic heavy ion collisions in ground laboratory. The initial  parton systems produced at Relativistic Heavy-Ion Collider (RHIC)
and Large Hadron Collider (LHC) have an anisotropic momentum distribution  and experience a complicated evolution to equilibrium  finally.  The transport coefficients strongly influence the evolution  of the produced hot and dense QCD matter.
The relativistic hydrodynamic simulations associated with a small shear viscosity rather successfully explain the flow data and transverse momentum spectra of ultimate hadron from  heavy ion collisions, for reviews please refer to Refs\cite{heinz09,teaney09,romatschke10,gale13,song}. Soon afterwards the viscous effects on the diverse aspects of the hot QGP, including photon and dilepton production\cite{lin08,jeon14},  heavy quarkonium dissolution\cite{guo1,guo2}, energy loss suffered by the fast parton traveling through the QGP \cite{dusling,calzetta,jiang15,sarkar18} and the induced wakes\cite{betz09,betz14,neufeld08,jiang11,jiang12}, the quark polarization\cite{huang11} and the chiral magnetic/vortical effects\cite{jiang,liao,li}, the dielectric properties of the QGP medium\cite{jiang10,jiang13,jiang16} and the dynamical evolution of QCD matter during the first-order  phase transition from hadronic matter to quark matter\cite{feng18}, have been addressed in recent years.

At the very early stage of the ultrarelativistic heavy ion collisions\cite{lappi,matsui} and in the so-called magnetic scenario for the QGP in the
evolution process near $T_c$  region\cite{liao1,liao2}, there are color-electric flux tubes in system containing the strong color-electric fields.
At the same time,  the very strong magnetic field will be generated perpendicular to the reaction plane in off-central
collisions\cite{kharzeev,mag2,tuchin}.
The study on the electromagnetic properties of the quark-gluon plasma has attracted increasing interest in recent years.

Electric conductivity $\sigma$ reflects the electromagnetic response
of the medium to the applied  electric field which is a kind of transport coefficient describing  the charge transport.
The electric conductivity enters into the hydrodynamic evolution equations for the QCD plasma produced in heavy ion collisions and dominates  the space-time evolution of energy-momentum\cite{matsui} and influences the quark-gluon chemical equilibration\cite{eskola}. In off-central ultrarelativistic heavy ion collisions, the electric conductivity determines the duration  of the  transient strong magnetic field   and the  strength of the chiral magnetic effect\cite{kharzeev,tuchin,fukushima}. As a result, the electric conductivity is responsible for the distribution of the charge-dependent flow of final state hadron sensitive to the chiral magnetic  and  separation effects in off-central heavy-ion collisions\cite{kharzeev,tuchin,fukushima,hirono}. Moreover, the electric conductivity significantly influences the soft photon and the low mass dilepton yield\cite{eskola,yin,ding83,ding}. In a word,
as shear viscosity, the electric conductivity plays an important role in  the matter evolution  in ultrarelativistic heavy ion collisions.

During the foundation of the QGP kinetic theory  in 1980s-1990s, several scientific groups have studied the electric conductivity of the QGP within that theoretical framework\cite{heinz85,heinz,czyz86,mrowczynski88,dyrek87,mrowczynski89,mrowczynski90}.
As the development of the thermal field theory and the QGP transport theory, people have investigated it with the resummation QCD perturbative theory and the effective kinetic theory subsequently \cite{selikhov93b,manuel04,hou,heiselberg94,arnold03,arnold99,bodeker98,bodeker99,litim99a,litim99b,blaizot99,blaizot00}. Recently, the electric conductivity of the QGP has  aroused people's attention  and has been studied with different approaches, including numerical solution of the Boltzmann equation \cite{puglisi14,greif}, the quasiparticle model \cite{puglisi15,srivastava17,mitra17}, the Dyson-Schwinger approach \cite{qin} and the lattice gauge theory \cite{ding83,ding}. In addition, the study of the electric conductivity in the QGP under magnetic field  has attracted increasing interest in heavy ion community \cite{fukushima18,feng17b,hattori16,hattori17,kurian,das}.

In a thermal plasma system, the temperature gradient in the spatial distribution and the velocity gradient between the adjacent fluid layers  will result in the heat transport and the momentum transport, respectively.  The heat conductivity $\lambda$ and the shear viscosity $\eta$ are the corresponding transport coefficients which dominate the system to approach the heat equilibrium and the momentum isotropy. An applied electric field will induce charge  transport  as well and the electric conductivity $\sigma$ characters the charge transport in the plasma system. Those  different  transport processes may take place at the same time in the QGP. In the present paper, we will study the electric conductivity of the QGP in presence of shear viscosity, which is focused on the mutual impacts between the different  transport phenomena in the QGP.

The relation between the heat transport and the charge transport  has been firstly addressed in condensed matter system in history. People have found that  in a metal the ratio of the heat conductivity over the electric conductivity timing the bulk temperature of the system $\frac{\lambda}{\sigma T}$ is a constant-Lorentz number, which is the so called Wiedemann-Franz law\cite{stat}.  Some scientific groups have investigated the Wiedemann-Franz law in the QGP and hot dense hadronic systems in Refs.\cite{mitra17,denicol19,sahoo19,rath19,rath} very recently. On the other hand, some investigations reported that  the ratio of shear viscosity over thermal conductivity $\frac{\eta c_p}{\lambda \rho}$ ($c_p$ and $\rho$ are the specific heat at constant pressure and mass density of the system) is also a  constant in dilute atomic Fermi gases\cite{braby} as well as in the QGP with finite chemical potential\cite{jaiswal15}, which is qualitatively similar to that obtained in a strongly coupled conformal plasma\cite{son06}. A constant result for   $\frac{\eta c_p}{\lambda \rho}$ is an interesting analogy to the Wiedemann-Franz law.

In the context of relativistic heavy ion collisions, the relation between shear viscosity and the electric conductivity  has been focused on recently in Refs.\cite{puglisi15,srivastava17,mitra17,sahoo,muller15}.
In those references, the authors have calculated the shear viscosity and the electric conductivity  separately under the same theoretical framework with the same physical conditions. Then, based on the derived results the ratio of the shear viscosity over the electric conductivity is performed to address
that which transport process will play a relatively significant role in the evolution of the QGP \cite{puglisi15,srivastava17,mitra17,sahoo,muller15}. In the present work, we will study the connection between the electric conductivity and the shear viscosity with an alternative approach. According to the Refs.\cite{lin08,dusling,groot,teaney,teaney08}, viscosity will modify the distribution functions of the constituents of the QGP.
Based on the chromohydrodynamic equations obtained from the QGP kinetic theory and the distribution function modified by shear viscosity, one can derive the induced color current through which conductivity tensor will be derived. Through the viscous distribution functions, viscous chromohydrodynamic equations, the induced color current and the conductivity tensor,  shear viscosity will embed into the electric conductivity.  Therefore, one can study the viscous effects on the  conductivity properties. Quarks carry not only electric charges, but also color charges. The dynamics of quarks is governed by quantum chromodynamics, thus we should call the research object the color-electric conductivity.

The paper is organized as follows. In section 2, we will briefly review the formalism for the calculation of the electric conductivity. In the next section, by solving the viscous chromohydrodynamic equations formulated from the QGP kinetic theory together with the distribution function modified by the shear viscosity, we will derive the induced color current through which  the conductivity tensor can be abstracted according to the linear response theory.
In section 4, we will evaluate the color-electric conductivity and study the viscous effects on it. Section 5 is summary.

The natural units $k_B=\hbar=c=1$, the metric $g_{\mu\nu}=(+,-,-,-)$ and the notations $k^\mu=(\omega,\mathbf{k})$ and $K^2=\omega^2-k^2$ are used in the paper.

\section{Color-electric conductivity in kinetic theory}\label{emprop}
The electric conductivity can be evaluated from the Kubo formula related to the current-current correlation for a system in thermal equilibrium. It also can be derived by extracting the proportionality coefficient of the  induced electric current responding to the applied external electric field  according to the Ohm's law $\textbf{j}=\sigma \textbf{E}$. Some people have extended the Ohm's law to a covariant form applicable to the non-Abelian plasma\cite{matsui,heinz85,heinz,czyz86,mrowczynski88,dyrek87,mrowczynski89,mrowczynski90,selikhov93b}
\begin{equation}\label{c1}
j^\mu_a=\sigma^{\mu\nu}_{ab}F^b_{\nu\lambda}u^\lambda,  \ \ \ \ \ \ \ a,b=1,2,...,8,
\end{equation}
where $F^b_{\nu\lambda}$ and $u^\lambda$ are the field strength tensor and the fluid velocity, respectively. The current $j^\mu_a$ is induced by the color-electric field $F^b_{\nu\lambda}u^\lambda$, therefore the proportionality coefficient $\sigma^{\mu\nu}_{ab}$ is color-electric conductivity tensor through which we can study the conductivity properties.

The kinetic equation for partons is given by~\cite{heinz,mrowczynski89,mrowczynski90,manuel04}
\begin{equation}
p^\mu D_\mu Q^i(p,x)+\frac{g}{2}\theta^ip^\mu
\{F_{\mu\nu}(x),\partial^\nu_pQ^i(p,x)\}=C,\label{kin}
\end{equation}
$Q^i(p,x)$ with $i\in{g,q,\overline{q}}$  denote the
distribution functions of gluon, quark and  antiquark,  respectively, which are $(N_c^2-1)\times(N_c^2-1)$  and $N_c\times N_c$  matrices respectively. $\theta^{g}=\theta^{q}=1$, $\theta^{\bar{q}}=-1$ and $\partial_{\nu}^{(p)}$ denotes the four-momentum derivative.
$D_\mu$ ($\mathcal{D}_\mu$) represents the covariant derivatives $D_\mu=\partial_\mu-ig[A_\mu(x),\cdots]$($\mathcal{D}_\mu=\partial_\mu-ig[\mathcal{A}_\mu(x)\cdots]$) with the gauge field
$A^{\mu}=A^{\mu}_a \tau^a$ ($\mathcal{A}^{\mu}=\mathcal{A}^{\mu}_a T^a$), where $\tau^a$($T^a$) is the $\rm SU(N_c)$ group generators in the fundamental(adjoint) representation with $\mathrm{Tr}[\tau^a,\tau^b]=\frac{1}{2}\delta^{ab}$($\mathrm{Tr[T^a,T^b]}=N_c\delta^{ab}$).
 $F_{\mu\nu}=\partial_\mu
A_\nu-\partial_\nu A_\mu-ig[A_\mu,A_\nu]$ represents the strength
tensor in the fundamental representation, and $\mathcal{F}_{\mu\nu}$
is its counterpart  in the adjoint representation.
$\emph{C}$ is the collision term.

The transport equations are supplemented by the Yang-Mills equation
$D_\mu F^{\mu\nu}(x)=j^\nu(x)$, and
the color current $j^\nu(x)$  is given  in the fundamental
representation as
\begin{eqnarray}\label{current}
j^\nu(x)=-\frac{g}{2}\int_pp^\nu[Q^q(p,x)-Q^{\bar{q}}(p,x)-\frac{1}
{3}Tr[Q^q(p,x)\\ \nonumber-Q^{\bar{q}}(p,x)]+2\tau^aTr[T^aQ^g(p,x)]],
\end{eqnarray}
where $\int_p=\int \frac{d^4p}{(2\pi)^3}2 \Theta(p_0) \delta(p^2).$

In the Vlasov approximation $C=0$, by solving the QGP transport equations (\ref{kin}) and (\ref{current}) associated with Eq.(\ref{c1}) in the linear response approximation,  one can obtain the color-electric conductivity
of the QGP\cite{heinz85,heinz,mrowczynski88,mrowczynski89,mrowczynski90} in the hard thermal loop (HTL) approximation.

\section{Color-electric Conductivity in Viscous Chromohydrodynamics}\label{na}

\subsection{Viscous chromohydrodynamic equations}\label{nb}

Viscosity will modify the distribution functions of the constituents of a microscopic system\cite{lin08,dusling,groot,teaney,teaney08}. If shear viscosity is taken into account only, the distribution function denotes as
\begin{equation}
Q=Q_o+\delta Q=Q_o+\frac{c'}{2T^3}\frac{\eta}{s}Q_o(1\pm Q_o)p^\mu
p^\nu\langle \nabla_\mu u_\nu \rangle. \label{dis}
\end{equation}
In Eq.(\ref{dis}),``$+$''(``$-$'') is for boson(fermion) and $c'=\pi^4/90\zeta(5)$ ($c'=14
\pi^4/1350\zeta(5)$) is for massless boson(fermion)~\cite{lin08,dusling,teaney,teaney08}.
$\langle
\nabla_\mu u_\nu \rangle = \nabla_\mu u_\nu + \nabla_\nu u_\mu -
\frac{2}{3} \Delta_{\mu\nu}\nabla_{\gamma}u^{\gamma}$, $\nabla_{\mu}
= (g_{\mu\nu} - u_{\mu}u_{\nu})\partial^{\nu}$, $\Delta^{\mu\nu}=
g^{\mu\nu} - u^{\mu}u^{\nu}$;   $\eta, s$,   $T$, $Q_o$
represent the  shear viscosity, the entropy density,  the temperature of the system and the equilibrium distribution function of boson or fermion, respectively. That ansatz of distribution function is widely used in hydrodynamic simulations to study phenomenology of relativistic heavy ion collisions.

It should be noted that only  when the system is slightly off-equilibrium, it is possible to evaluate the small departure of the distribution $\delta Q$ from its equilibrium value $Q_o$ due to non-equilibrium effect\cite{groot}. The viscous corrected distribution function used in (\ref{dis}) is much smaller than the equilibrium one $\delta Q<<Q_o$, which implies a small value of the velocity gradient.
Therefore, the color-electromagnetic fields dominate interaction in QGP system and the collision terms in transport equations may be neglected.
In Refs.\cite{jiang10,jiang13}, the authors have extended the ideal chromohydrodynamic equations\cite{mrowczynski90,manuel06,manuel07} to the viscous ones by expanding the collisionless kinetic equation (\ref{kin}) in momentum moments in terms of distribution function modified by shear viscosity (\ref{dis}). It is argued that chromohydrodynamics can describe the polarization effect as the kinetic theory\cite{jiang13,manuel08}. In addition,
the fluid equations dealing with conservative equations of the macroscopic physical quantities are much simpler than those of the kinetic theory. Therefore, one can study the connection of the color-electric conductivity and shear viscosity of a quark gluon system with viscous chromohydrodynamics.

The constitutive equations for the viscous chromohydrodynamics are~\cite{jiang10,jiang13},
\begin{equation}
D_\mu n^\mu=0, \ \ \ \ \   D_\mu
T^{\mu\nu}-\frac{g}{2}\{F^\nu_\mu,n^\mu(x)\}=0   \label{con1}
\end{equation}
with
\begin{eqnarray}
 n^\mu(x)=\int_p  p^\mu Q(p,x), \
\    T^{\mu\nu}(x)=\int_p  p^\mu p^\nu Q(p,x), \label{co}
\end{eqnarray}
where $Q(p,x)$ is the distribution function modified by shear viscosity (\ref{dis}).
For detailed derivation of viscous chromohydrodynamic equations, please refer to Refs.\cite{jiang10,jiang13,manuel07}.

The four-flow $n^\mu$ and energy momentum tensor $T^{\mu\nu}$  can be expressed in the form\cite{jiang10,jiang13}
\begin{eqnarray}
n^\mu=n(x) u^\mu, \ \ \ \ \ \ \ \ \ \ \ \ \ \ \ \ \ \  \ \ \ \ \ \ \ \nonumber\\
T^{\mu\nu}=\frac{1}{2}(\epsilon(x)+p(x))
\{u^\mu,u^\nu\}-p(x)g^{\mu\nu}+\pi^{\mu\nu}, \label{num}
\end{eqnarray}
where
\begin{eqnarray}
\pi^{\mu\nu}=\eta \langle \nabla^\mu u^\nu \rangle = \eta
\{(g^{\mu\rho}-u^\mu u^\rho)\partial_\rho u^\nu+(g^{\nu\rho}\\\nonumber-u^\nu
u^\rho)\partial_\rho u^\mu-\frac{2}{3}(g^{\mu\nu}-u^\mu
u^\nu)\partial_\sigma u^\sigma\}.\label{vis}
\end{eqnarray}
If $\eta=0$, the distribution function (\ref{dis})
remains the ideal form, $\pi^{\mu\nu}$ will be absent in (\ref{num})
and the chromohydrodynamic equations will turn to the ideal ones\cite{manuel06,manuel07}.

The color
current (\ref{current})  reads
\begin{equation}
j^\mu(x)=-\frac{g}{2}(nu^\mu-\frac{1}{3}Tr[nu^\mu]).\label{curr}
\end{equation}
Eqs.(\ref{con1}),(\ref{num}) and (\ref{curr}) make up the basic set of equations of the viscous chromohydrodynamics.
In those equations,  $n$, $\epsilon$ and $p$ represent  the particle density, the energy density and pressure respectively.  Usually, hydrodynamic quantities have both colorless and colorful parts, as an example,
the particle density can be written as\cite{jiang10,jiang13,manuel07}
\begin{equation}
n^\mu_{\alpha\beta}=n_0^\mu I_{\alpha\beta}+\frac{1}{2}n_a^\mu \tau^a_{\alpha\beta}\label{color}
\end{equation}
where $\alpha,\beta=1,2,3$ are color indices and $I$ is the identity
matrix \cite{manuel07}.

\subsection{Induced color current}

To perform further analysis, we will linearize the hydrodynamic quantities around the stationary,
colorless and homogeneous state which is described by
$\bar{n}$,$\bar{u}^\mu$,$\bar{p}$ and $\bar{\epsilon}$. As an
example, the particle density $n(x)$ can be denoted  as
\begin{equation}\label{linearization}
n(x)=\bar{n}+\delta n(x).
\end{equation}
The covariant derivatives of the hydrodynamic quantities in the stationary,
colorless and homogeneous state vanish, for example $D_\mu\bar{n}=0$.   In that state the
color current $j^\mu(x)=0$.
The diagonalized fluctuation quantity should be much smaller than  the corresponding stationary one $\delta n\ll \bar{n}$\cite{jiang10,jiang13,manuel07}.   As Eq.(\ref{color}) all  fluctuation quantities  contain both
colorless and colorful components,
\begin{equation}\label{linearization1}
\delta n_{\alpha\beta}=\delta n_0 I_{\alpha\beta}+\frac{1}{2}\delta n_a
\tau^a_{\alpha\beta}.
\end{equation}

Substituting the linearized hydrodynamic quantities like
Eqs.(\ref{linearization})(\ref{linearization1}) into Eq.(\ref{num}) and their corresponding
conservation equations (\ref{con1}) and projecting them on
$\bar{u}^\mu$ and
$(g^{\mu\nu}-\bar{u}^\mu\bar{u}^\nu)$,
then, considering only the equations for colorful parts of fluctuations and performing
the Fourier transformation, one can obtain equations which can
describe color phenomena in the viscous QGP\cite{jiang10,jiang13}
\begin{equation}
\bar{n}k_\mu\delta u^\mu_a+ k_\mu\delta
n_a\bar{u}^\mu=0,\label{con1f}
\end{equation}
\begin{equation}
\bar{u}^\mu
k_\mu\delta\epsilon_a+(\bar{\epsilon}+\bar{p})k_\mu\delta
u^\mu_a=0,\label{con21f}
\end{equation}
\begin{eqnarray}\label{con22f}
(\bar{\epsilon}+\bar{p})(\bar{u}\cdot K)\delta
u^\nu_a+(-k^\nu+\bar{u}^\nu(\bar{u}\cdot K))\delta p_a + \eta\{(K^2
\\\nonumber-(K\cdot \bar{u}))\delta u^\nu_a  + (k^\mu k^\nu-k^\mu
\bar{u}^\nu)\delta u_{\mu,a}+\frac{2}{3}(\bar{u}^\nu (K\cdot
\bar{u})\\ \nonumber-k^\nu)k_\rho\delta u^\rho_a\}=ig\bar{n}\bar{u}_\mu
F_a^{\mu\nu}(K),
\end{eqnarray}
with $\bar{u}\cdot K=\bar{u}^\mu k_\mu$.

According to Eq.(\ref{curr}), the
color current due to the color fluctuations of the hydrodynamic quantities  is given by\cite{jiang10,jiang13}
\begin{equation}\label{curr1}
j^\mu_a=-\frac{g}{2}(\bar{n}\delta u^\nu_a+ \delta
n_a\bar{u}^\mu-\frac{1}{3} \rm{Tr} [\bar{n}\delta u^\mu_a+ \delta n_a\bar{u}^\mu]).
\end{equation}

By introducing an EoS $\delta p_a=c_s^2\delta\epsilon_a$ to complete the fluid equations (\ref{con1f})(\ref{con21f})(\ref{con22f}) (the explicit formulism for $c_s$ will be introduced later),
we can solve the color fluctuations of
hydrodynamic quantities $\delta n_a$, $\delta u_{\nu,a}$ and
$\delta\epsilon_a$\cite{jiang10,jiang13},
\begin{equation}
\delta n_a=-\frac{\overline{n}k_\mu\delta u_a^\mu}{K\cdot
\overline{u}}, \ \ \ \ \
\delta\epsilon_a=-\frac{(\overline{\epsilon}+ \overline{p})k_\mu\delta
u_a^\mu}{K\cdot \overline{u}},  \label{reso}
\end{equation}
\begin{eqnarray}
\delta u_{\sigma,a}=\frac{1}{1+D(K^2-(K\cdot \overline{u})^2)}\cdot
\frac{g\overline{n}}{(\overline{\epsilon}+\overline{p})(K\cdot\overline{
u})}\{g_{\sigma\nu}\\\nonumber+(B+E)(k_\sigma k_\nu-\overline{u}_\sigma
k_\nu(K\cdot \overline{u}))\}\cdot \overline{u}_\mu iF_a^{\mu\nu},
\label{reso2}
\end{eqnarray}
with
\begin{eqnarray}\label{d}
 B&=&-\frac{c_s^2}
  {\omega^2-c_s^2k^2}, \ \ \ \ \ \ \ \ D=\frac{\eta}{sT\omega}, \\\nonumber
  E&=&-\frac{\frac{\eta\omega}{sT}
 (1+4\frac{c_s^2k^2}
  {\omega^2-c_s^2k^2})}
  {3\omega^2-3c_s^2k^2-
 4\frac{\eta\omega k^2}
{sT}}.
\end{eqnarray}

Substituting  the solved $\delta n_a$ and $\delta u^\mu_a$ into Eq.(\ref{curr1}),
one can get the induced color current due to the color fluctuations of hydrodynamic quantities
\begin{eqnarray}\label{currf}
 j_a^\mu
=-\frac{i\omega^2_p}{(K\cdot\overline{
u})}\cdot \frac{1}{1+D(K^2-(K\cdot \overline{u})^2)}(g^{\mu\sigma}-\frac{\overline{u}^\mu k^\sigma}
 {K\cdot \overline{u}}) \\\nonumber
\cdot \{g_{\sigma\nu}+(B+E)(k_\sigma k_\nu-\overline{u}_\sigma
k_\nu(K\cdot \overline{u}))\}\overline{u}_\rho F_a^{\rho\nu}
\end{eqnarray}
where $\omega^2_p=\frac{g^2\bar{n}^2}{2(\bar{\epsilon}+\bar{p})}$ is the  square of the plasma frequency.
For detailed derivation of $j_a^\mu$ in chromohydrodynamic approach, please refer to the Refs.\cite{jiang10,jiang13}.

\subsection{Color-electric conductivity}

According to Eq (\ref{c1}), we can extract the conductivity tensor from equation (\ref{currf})
\begin{eqnarray}\label{cten}
 \sigma_{ab}^{\mu \nu}=
-\frac{i}{\omega}\frac{\delta^{ab}\omega_p^2}{1-Dk^2}\{g^{\mu \nu}+(B+E)(k^\mu k^\nu-\overline{u}^\mu k^\nu (K\cdot \overline{u}) )
\\\nonumber-\frac{\overline{u}^\mu k^\nu}{(K\cdot \overline{u})}-(B+E)(\frac{K^2\overline{u}^\mu k^\nu}{K\cdot \overline{u}}-\overline{u}^\mu k^\nu (K\cdot \overline{u})))
\}.
\end{eqnarray}
The diagonalized spatial component of $\sigma_{ab}^{\mu \nu}$  in the color space reads from equation (\ref{cten}),
\begin{eqnarray}\label{conf}
 \sigma^{ij}=-\frac{i}{\omega} \frac{\omega_p^2}{1-Dk^2} \{g^{ij}+(B+E)k^ik^j
 \}.
\end{eqnarray}


For an isotropic, homogeneous plasma medium, $\sigma^{ij}$ reduces to two scalar functions according to project operators\cite{heinz,blaizot00,dressel}
\begin{eqnarray}\label{decomp}
 \sigma^{ij}=\sigma_L\frac{k^ik^j}{k^2}+\sigma_T(\delta^{ij}-\frac{k^ik^j}{k^2}).
\end{eqnarray}
$\sigma_L$ and $\sigma_T$ are the longitudinal and transverse color-electric conductivities which are independent  response functions and do not mix.  From the relation between the dielectric functions and the conductivities,  the longitudinal conductivity describes the response of the medium to a scalar potential $\phi$ and the transverse one reflects the medium response to a vector potential $\textbf{A}$ in electrodynamics, for details please refer to $\S 3.1$ in Ref.\cite{dressel}.

According to Eqs.(\ref{conf}) and (\ref{decomp}), we can obtain
\begin{eqnarray}\label{cl}
 \sigma_L=\sigma^{ij}\frac{k_ik_j}{k^2}=\frac{i}{\omega} \frac{\omega_p^2}{1-Dk^2}\{1-(B+E)k^2
 \},
\end{eqnarray}
and
\begin{eqnarray}\label{ct}
 \sigma_T=\frac{1}{2}(\delta_{ij}-\frac{k_ik_j}{k^2})\sigma^{ij}=\frac{i}{\omega} \frac{\omega_p^2}{1-Dk^2}.
\end{eqnarray}

\begin{figure}
\centering{\includegraphics{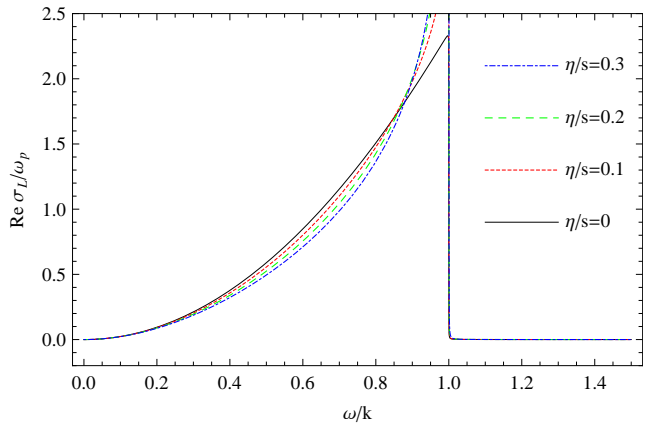}}
\caption{The  real part of  the scaled longitudinal color-electric conductivity of the QGP for different ratios of shear viscosity over entropy density $\frac{\eta}{s}=0,0.1,0.2,0.3$, respectively.}\label{light}
\end{figure}

Substituting $B$, $D$ and  $E$ mentioned in (\ref{d}) into (\ref{cl})(\ref{ct}) and adopting the effective sound speed
$c_s=\sqrt{\frac{1}{3(1+\frac{1}{2y}\log\frac{1-y}{1+y})}+\frac{1}{y^2}}$
 $(y=\frac{k}{\omega})$\cite{jiang10,jiang13,manuel06,manuel07}, one can obtain the scaled longitudinal  color-electric conductivity
\begin{eqnarray}\label{conr}
\frac{\sigma_L(\omega,k)}{\omega_p}
 &=&-\frac{i\omega}{1-\frac{\eta k^2}
 {sT\omega}}\frac{3\omega_p}{k^2}(1-\frac{\omega}{2k}
 \log\frac{\omega+k+i\xi}{\omega-k+i\xi})\\\nonumber&+&\frac{i\omega_p}
 {1-\frac{\eta k^2}{sT\omega}}
 \frac{\eta}{sT}\frac{1}
  {1+ 4\frac{\eta\omega }
{sT}{(1-\frac{\omega}{2k}\log\frac{\omega+k+i\xi}{\omega-k+i\xi})}}\\ \nonumber&\cdot& \{3(1-\frac{\omega}{2k}
 \log\frac{\omega+k+i\xi}{\omega-k+i\xi})\\\nonumber&+&\frac{12\omega^2}{k^2}(1-\frac
 {\omega}{2k}
\cdot\log\frac{\omega+k+i\xi}{\omega-k+i\xi})^2\},
\end{eqnarray}
and the transverse one
\begin{equation}\label{contr}
 \frac{\sigma_T}{\omega_p}=\frac{i\omega_p}{\omega} \frac{1}{1-\frac{\eta k^2}{s\omega T}}.
\end{equation}
As shown in Eqs.(\ref{conr})(\ref{contr}), the longitudinal and transverse conductivities  are usually complex functions of $\omega,k$. It is argued that in electrodynamics the real part of the conductivity describes a finite dissipation of energy\cite{dressel}. While the imaginary part of the conductivity
defines the phase lag between the applied electric field and the induced electric current which manifests that the medium response has a time delay with respect to the applied disturbation\cite{dressel,yang15}.

\section{Results and discussion}
\label{summary}

\begin{figure}
\centering{\includegraphics{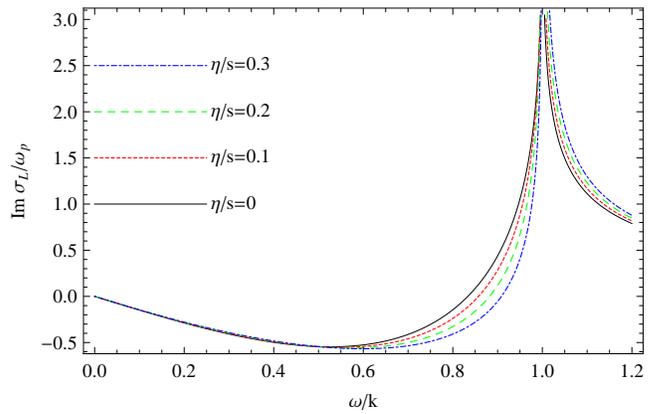}}
\caption{The  imaginary part of the scaled longitudinal color-electric conductivity of the QGP for different ratios of shear viscosity over entropy density $\frac{\eta}{s}=0,0.1,0.2,0.3$, respectively.}\label{charm}
\end{figure}

As the collisionless approximation of the QGP transport equations are adopted in the derivation of the chromohydrodynamic equations, we may obtain the conductivities  in the level of HTL approximation in the presence of shear viscosity. When $\eta/s=0$, (\ref{conr}) turns to
\begin{eqnarray}\label{e0}
\sigma_L(\omega,k)
 =-\frac{3i\omega\omega_p^2}{k^2}(1-\frac{\omega}{2k}
 \log\frac{\omega+k+i\xi}{\omega-k+i\xi}).
\end{eqnarray}
One  can  obtain (\ref{e0}) in terms of the dielectric tensor $\varepsilon^{ij}$ in the HTL approximation\cite{heinz,mrowczynski89,mrowczynski90,blaizot00} and  (\ref{decomp})(\ref{cl}) associated with the relation between the conductivity tensor and the dielectric tensor (\ref{sigl})\cite{blaizot00,mrowczynski88,mrowczynski89,mrowczynski90,dressel} discussed in the following . Therefore, (\ref{e0}) is just the HTL result of the longitudinal color-electric conductivity. According to (\ref{e0}), one can get the  real part of $\sigma_L$
\begin{eqnarray}\label{ehtl}
Re \sigma_L(\omega,k)
 =\frac{3\pi\omega_p^2}{4k}\frac{2\omega^2}{k^2}\Theta(k^2-\omega^2).
\end{eqnarray}
$\Theta$ is the Heaviside step function which equals to unit for $k>\omega$.
There is  a discrepancy of the factor $\frac{2\omega^2}{k^2}$ between
(\ref{ehtl}) and (3.36) in Ref.\cite{arnold99} obtained from the transverse polarization self-energy $\Pi_T(\omega,k)$ within the framework of
the Boltzmann equation in the collisionless limit and $k>>\omega$.


As shown in Eq.(\ref{contr}), the transverse color-electric conductivity $\sigma_T$ is a pure imaginary function of $\omega,k$. When $\eta/s=0$,
(\ref{contr}) can not recover the HTL result  which can be derived in terms of the HTL dielectric tensor and Eqs.(\ref{sigl})(\ref{decomp})(\ref{ct}). The viscous chromohydrodynamics can describe polarization effect as the kinetic theory,  which facilitate us to investigate the viscous effect on the electromagnetic properties of the QGP.
However, some dynamical information will be lost during the derivation from the kinetic theory to the chromohydrodynamics\cite{manuel06,manuel07,manuel08,jiang15}.
Nevertheless, the chromohydrodynamics can describe  some dynamical information of system relevant to the longitudinal dielectric properties of the plasma, which could still capture some interesting physics of the QGP.


From discussion in last two paragraphs in this section, it indicates that in chromohydrodynamic approach the conductivity properties of the QGP are mainly demonstrated  by the longitudinal part of the color-electric conductivity $\sigma_L(\omega,k)$. It should be noted that another study shows that the conductivity properties of the QGP under the magnetic field also mainly come from the longitudinal part. To the one-loop order, the transverse (with respect to the magnetic field direction) electric conductivity of the QGP  vanishes $\sigma_\bot=0$\cite{fukushima18}.

In the following, we will apply  $k=0.2\omega_p$ and $T=\omega_p$ to perform numerical analysis to study
$\omega$-dependent behavior of the color-electric conductivity with different values of shear viscosity.
We present the real  part of the scaled longitudinal color-electric conductivity with respect to frequency with different shear viscosity in Fig.~1.  The black, red, green and blue curves are the cases of $\eta/s=0,0.1,0.2,0.3$ respectively.
One can see from  Fig.1 that in the space-like region $\frac{\omega}{k}<1$, the real part of the longitudinal color-electric conductivity increases with the frequency monotonously.  At a small value of frequency $\frac{\omega}{k}\leq 0.3$, the viscous curves of $\rm{Re} \sigma_L/\omega_p$ superpose each other nearly, shear viscosity has a trivial effect on the  conductivity.  As the frequency increases, the viscous effects on the color-electric conductivity become appreciable and $\rm{Re} \sigma_L/\omega_p$ reduces as the increase of shear viscosity. While for $\frac{\omega}{k}>0.875$,  for a fixed frequency, the larger the shear viscosity, the larger the value of $\rm{Re} \sigma_L/\omega_p$, which shows a reversed dependence of the longitudinal color-electric conductivity on shear viscosity  compared to the case in the small frequency region.
In the time-like region $\frac{\omega}{k}>1$, the real part of the longitudinal color-electric conductivity turns to vanish.

The imaginary part of the longitudinal color-electric conductivity is displayed in Fig.2.  In the space-like region there is a critical frequency which is aroud $\omega_c\sim 0.6k$, the imaginary part of the longitudinal color-electric conductivity  decreases with frequency  for $\omega\leq\omega_c$. It indicates that shear viscosity has no demonstrable effects on the imaginary part of the  conductivity in that frequency region. For $\omega>\omega_c$, $\rm{Im} \sigma_L/\omega_p$ rises quickly as the increase of frequency. At the same time, viscous effects on $\rm{Im} \sigma_L/\omega_p$ become remarkable and $\rm{Im} \sigma_L/\omega_p$  diminishes as shear viscosity increases.  In the time-like region, $\rm{Im} \sigma_L/\omega_p$ reduces with frequency, while shear viscosity enhances the imaginary part of the longitudinal color-electric conductivity.

From Eq.(\ref{conr}), it is clear that the real part  of  $\sigma_L/\omega_p$ is determined by the imaginary part  of the logarithmic function $\log \frac{\omega+k+i\xi}{\omega-k+i\xi}$, and $\rm{Im}\sigma_L/\omega_p$ is related to the real part of that  function correspondingly. The logarithmic function can be expressed by\cite{jiang10,mrowczynski90}
\begin{equation}\label{logfun}
\log \frac{\omega+k+i\xi}{\omega-k+i\xi}=\log\mid \frac{\omega+k}{\omega-k}\mid-i\pi\Theta(k^2-\omega^2),
\end{equation}
In the time-like region $\omega>k$, $\Theta(k^2-\omega^2)=0$ which results in $\rm{Re} \sigma_L/\omega_p$ turning  to vanish in that frequency region, as shown in Fig.1. There is a singularity located at the position $\omega=k$ for  the real part of the logarithmic function $\log\mid \frac{\omega+k}{\omega-k}\mid$, which leads to a divergence behavior of $\rm{Im} \sigma_L/\omega_p$ at $\omega/k=1$ as shown in Fig.2.

It is instructive to achieve an understanding in viscous effects on the color-electric conductivity from the view of point of dielectric properties.
The fields in medium are different from those in vacuum and the dielectric functions dominate those differences.
Moreover, other electromagnetic properties of the medium   can be derived from the latter in principle.
Mrowczynski et al have  found that the dielectric tensor and the conductivity tensor are related with each other as following  \cite{mrowczynski88,mrowczynski89,mrowczynski90}
\begin{equation}\label{sigl}
 \sigma_{ab}^{\alpha\beta}(k)=-i\omega[\varepsilon_{ab}^{\alpha\beta}(k)-\delta^{\alpha\beta}\delta_{ab}].
\end{equation}
If the dielectric tensor and the conductivity tensor are expressed according to project operators (\ref{decomp}) simultaneously, one can obtain
\begin{equation}\label{condie}
 \sigma_L=
 -i\omega\{\varepsilon_L-1 \}.
\end{equation}
In (\ref{condie}), $\varepsilon_L$ is the longitudinal dielectric function. So from (\ref{condie}), the real part   of the longitudinal color-electric conductivity  relates to the imaginary part  of the longitudinal dielectric function. And one can obtain the imaginary part of $\sigma_L$  from the real part of $\varepsilon_L$.
The longitudinal color-electric conductivity (\ref{conr}) is consistent with  the result of the longitudinal dielectric function obtained in Refs.\cite{jiang10,jiang13} in terms of Eq.(\ref{condie}). Furthermore, if the HTL dielectric function $\varepsilon_L$\cite{heinz85,heinz,mrowczynski89,mrowczynski90}  is applied to (\ref{condie}), we will obtain (\ref{e0}), \ie the result of the longitudinal conductivity in the HTL approximation.

It is argued in Ref.\cite{jiang} that the electric conductivity arises from competition between ``ordered'' electric force and  ``disordered'' scatterings. Thus the dissipation involving the disordered scatterings will affect the charge transport.
The electric conductivity reflects the medium response to the applied external electric field. The medium properties may be involved in its response to the applied fields.
Since shear viscosity  modifies the distribution functions of the medium constituents, which may play an important role in determining the medium response. Therefore shear viscosity will have an impact on the charge transport naturally.

It should be stressed that when several different irreversible transport processes (such as heat conductivity, electric conductivity and shear viscosity) take place at the same time in a plasma medium the transport processes may interfere with each other\cite{onsager1}.  The driving force due to the gradient of a kind of physical quantity can result in another kind current\cite{onsager1}. For example,  an electric current will produce  in a circuit composed of different metals  when the junctions are at different temperatures, which is known as thermoelectric effect or Seebeck effect\cite{stat}. Some authors have addressed the Seebeck effect  for the hot and dense hadronic matter with a temperature gradient\cite{bhatt2019} and the QGP in magnetic fields\cite{zhang20} in the context of ultra-relativistic heavy ion collisions. Recently, another study shows that a density gradient of a given charge can generate dissipative currents of another charges\cite{greif18}.
Therefore, it is natural to expect that the  velocity gradient might produce electric current in the system. As a result shear viscosity could influence the  charge transport coefficient, \ie the electric conductivity.

With viscous chromohydrodynamic approach we centred our attention  on physics related to the color fluctuation phenomena in QGP at a short time scale where the collision terms of transport equations can be neglected\cite{manuel06,manuel07}.
The collisionless kinetic theory has usually been applied to study the QGP properties  and some important results have achieved which  are coincided with those derived from the diagrammatic  approach with hard loop approximation---the  HTL approximation.
We also applied the collisionless transport equations to derive chromohydrodynamic equations but incorporating dissipative effects.
The neglect of the collision terms in transport theory does not imply that there are no dissipative interaction in plasma. It is argued in Ref.\cite{jiang13} that besides the collision terms, interaction between particles due to the mean field\cite{balescu} and the turbulent plasma fields\cite{asakawa1,asakawa2} also can induce dissipation in plasma. Besides the neglect of the collision terms, the derivation from the kinetic theory to chromohydrodynamics also results in the loss of some dynamical information\cite{manuel06,jiang13}. Nevertheless, the fluid equations still have rich dynamical content, which could capture some correct physics of the QGP. Therefore, we expect that we could gain some insight into the physics of the electromagnetic properties of the QGP by applying the viscous chromohydrodynamics.

\section{Summary}

In the linear response approximation, we solved  the viscous chromohydrodynamic equations which are derived from the QGP kinetic theory associated with the  distribution function modified by shear viscosity. According to the solved color fluctuations of the hydrodynamic quantities, we obtained the induced color current through which the conductivity tensor can be derived. Through the distribution function, viscous chromohydrodynamic equations and the fluctuating quantities of fluid, shear viscosity encodes in the induced color current and the conductivity tensor and  gives a corrective contribution to the color current and the color-electric conductivity. Generally the corrective color current due to shear viscosity  is much smaller than that induced by the applied external field.  Nevertheless, shear viscosity has an appreciable effect  on the color-electric conductivity in some frequency region.
Numerical analysis indicates that the conductivity properties  of the QGP are mainly demonstrated  by the longitudinal part of the color-electric conductivity.
In the space-like region, for a small frequency, shear viscosity has a trivial effect on the real and imaginary parts of $\sigma_L/\omega_p$. As the increase of frequency, viscous effects become notable and shear viscosity reduces both the real and imaginary parts of the conductivity in most of the space-like region. In the time-like region, the real part of the longitudinal conductivity turns to vanish, while shear viscosity increases the imaginary part of $\sigma_L/\omega_p$.

In the early stage of ultrarelativistic heavy ion collisions, the produced strongly interacting matter will have a large temperature gradient between the central and peripheral regions of the fireball and have an anisotropic momentum distribution between longitudinal  and transverse expansion. At the meanwhile, strong electromagnetic fields will produce in noncentral collisions. Therefore, there will exist several different transport processes simultaneously in the produced parton system in heavy ion collisions
which will result in interferences between different transports  in the system.
It is argued that the coupling of the bulk viscous pressure to shear-stress tensor (shear-bulk coupling)
can give an extra contribution to  the bulk viscous pressure. The extra part of the bulk viscous pressure due to shear-bulk coupling can be  comparable to the one originating from the Navier-Stokes term, which will remarkably affect the evolution process of the  QCD plasma produced in heavy ion collisions\cite{denicol14a,denicol14b,jaiswal,bazow}. Therefore, one can expect that the connection between shear viscosity and electric conductivity may be relevant to some observables in
ultrarelativistic heavy-ion collisions, which may be an
attractive issue and deserves a further comprehensive
investigation.

\begin{acknowledgments}
We thank Yun Guo, Yu-xin Liu, En-ke Wang and Peng-fei Zhuang for their instructive discussions.
The work is partially supported  by the NSFC Grant Nos.11465007, 11735007, 11890710, 11890711 and 11703005.
\end{acknowledgments}


\end{CJK*}

\end{document}